\documentclass[a4paper,11pt]{article}
\pdfoutput=1
\usepackage{jcappub}
\usepackage{ae,aecompl}
\usepackage{graphicx}
\usepackage{amsmath}
\usepackage{amssymb}
\usepackage{txfonts}
\usepackage{bm}
\usepackage{cleveref}
\usepackage{enumerate}
\def\dd{\mathrm{d}}

\def\equationautorefname~#1\null{%
  eq.~(#1)\null
}
\def\figureautorefname~#1\null{%
  figure~#1\null
}
\def\tableautorefname~#1\null{%
  table~#1\null
}

\newcommand{\ltsima}{$\; \buildrel < \over \sim \;$}
\newcommand{\lsim}{\lower.5ex\hbox{\ltsima}}
\newcommand{\gtsima}{$\; \buildrel > \over \sim \;$}
\newcommand{\gsim}{\lower.5ex\hbox{\gtsima}}
\newcommand{\bra}{\langle}
\newcommand{\ket}{\rangle}

\newcommand{\de}{\mathrm{d}}

\newcommand{\bs}[1]{\boldsymbol{#1}}

\title{\boldmath
Realistic systematic biases induced by residual intrinsic alignments in cosmic shear surveys
}
\author[a]{Robert Reischke}
\author[b]{and Bj\"orn Malte Sch\"afer}

\affiliation[a]{Department of Physics, Israel Institute of Technology -- Technion, 3200003 Haifa, Israel}
\affiliation[b]{Astronomisches Rechen-Institut, Zentrum f{\"u}r Astronomie der Universit{\"a}t Heidelberg,\\ Philosophenweg 12, 69120 Heielberg, Germany}
\emailAdd{r.reischke@campus.technion.ac.il}

\abstract{We study the parameter estimation bias induced by intrinsic alignments on a \texttt{Euclid}-like weak lensing survey. For the intrinsic alignment signal we assume a composite alignment model for elliptical and spiral galaxies using tidal shearing and tidal torquing as the alignment generating mechanism, respectively. The parameter estimation bias is carried out analytically with a Gaussian bias model and through running Monte-Carlo-Markov-chains on synthetic data including the alignment signal with a likelihood only including the cosmic shear signal. In particular, we study the impact of $II$ and $GI$ alignment terms individually as well as the more realistic situation where both types of alignment are present, and investigate the scaling of the estimation biases with varying strength of the alignment signal. Our results show that intrinsic alignments can cause substantial biases in cosmological parameters even if the coupling of galaxies to the ambient large is small. Especially $GI$-contributions strongly bias key cosmological parameters such as the dark energy equation of state. We also correct the analytic expression for the Gaussian bias model and find that the biases induced by intrinsic alignments are not accurately recovered by the simple analytic model.
}


\begin{document}
\maketitle
\flushbottom

\section{Introduction}
\label{sec:intro}
Measuring the weak gravitational lensing effect of the large-scale structure (LSS) \citep[for reviews see][]{bartelmann_weak_2001,hoekstra_weak_2008,kilbinger_cosmology_2015} is one of the primal science goals of upcoming large scale galaxy surveys such as \texttt{Euclid} or \texttt{LSST}. Given the amount of data delivered by these experiments the wealth of cosmological information accessible is unprecedented. To extract the lensing signal one has to assume that the shapes of background galaxies, acting as sources, are intrinsically uncorrelated \citep{kaiser_weak_1992}. This assumption, however, is not necessarily correct as galaxies interact with their surrounding which is imprinted in their shapes. Because physically close galaxies interact with the same patch of the cosmological large-scale structure, \citep[e.g.][]{schaefer_review:_2009,kiessling_galaxy_2015,troxel_intrinsic_2015,joachimi_intrinsic_2013} one is able to observed induced shape correlations, called intrinsic alignments (IA), which can be misinterpreted as an alignment signal \citep[see e.b.][for reviews]{kirk_galaxy_2015,kiessling_galaxy_2015,joachimi_intrinsic_2013,schaefer_review:_2009,troxel_intrinsic_2015}.

In general the alignment process of galaxies in the LSS is a complicated non-linear process which depends on galaxy formation and evolution. Some process can be made by tidal alignment models which link the shapes of galaxies to properties of the ambient LSS, very similar to local bias models \citep[for a detailed review see][]{desjacques_large-scale_2018} for galaxy clustering. The tidal alignment models split into a quadratic and a linear model for spiral and elliptical galaxies, respectively. For the latter, the gravitational tidal field distorts the isophotes of an elliptical galaxy by effectively perturbing the solution to the Jeans-equation \citep{hirata_intrinsic_2004,hirata_intrinsic_2010,blazek_tidal_2015}. The quadratic model relates the ellipticity to the angular momenta of spiral galaxies. Since the acquisition of angular momentum by spiral galaxies can be described by the misalignment between the inertial tensor and the tidal tensor, angular momenta of neighbouring galaxies are correlated and thus their shapes \citep{crittenden_spin-induced_2001,natarajan_angular_2001,mackey_theoretical_2002}. These models have been used extensively in the literature to model the effect of IA on weak lensing surveys \citep{blazek_tidal_2015,merkel_intrinsic_2013,tugendhat_angular_2017,krause_impact_2016,capranico_intrinsic_2013}. Apart from these models also more complicated approaches exist which try to take into account different aspects of galaxy formation \citep[e.g.][]{tenneti_intrinsic_2015,tenneti_galaxy_2014,codis_intrinsic_2015,chisari_intrinsic_2015} or for accounting for more complicated shapes of the local gravitational potential \citep{blazek_beyond_2017}. 

With the large uncertainty in the alignment process many studies try to remove the IA alignment signal from the analysis using different techniques which range from self-calibration, exploiting redshift dependencies, removing close pairs of galaxies, using multiple tracers or using the physical properties of the alignment mechanism \citep{zhang_proposal_2010,troxel_self-calibration_2012,huterer_uncorrelated_2005,joachimi_intrinsic_2010,shi_controlling_2010,crittenden_discriminating_2002,king_separating_2003,heymans_weak_2003,semboloni_effect_2013,merkel_theoretical_2014,munshi_tomography_2014,petri_sample_2016,hall_intrinsic_2014,larsen_intrinsic_2016,tugendhat_statistical_2018}.

In this paper we tend to investigate the impact of residual IA in cosmic shear maps on parameter inference. To this end we use the IA model presented in \citep{tugendhat_angular_2017} we then use analytic and Monte-Carlo-Markov-chains (MCMC) to estimate the resulting parameter estimation bias as a function of the residual alignment signal. While \citep{schaefer_angular_2015,tugendhat_angular_2017} already investigated the analytic bias we revisit the method used and correct the expression. We aim at an understanding to which degree IA models must be able to predict the IA signal in order to achieve scientific goals set for LSS experiments, especially we are looking for accurate and note only precise cosmological signals.

Throughout we will work with a $w$CDM model with the following fiducial values: $\Omega_\mathrm{m} = 0.32$, $\sigma_8 = 0.83$, $h=0.7$, $n_\mathrm{s} = 0.96$, $w_0 =-1$ and $w_\mathrm{a} = 0$. The paper is structured as follows: we review the shape correlations in \cref{sec:shape_correlations}. In \cref{sec:stat} the necessary statistical tools are presented. \Cref{sec:results} contains the results which are discussed and summarized in \cref{sec:conclusions}.

\section{Shape correlations}
\label{sec:shape_correlations}
All weak lensing studies rely on the observation of correlated ellipticities of background galaxies. To lowest order the observed ellipticity, $\epsilon$, is given by
\begin{equation}
\label{eq:observed_ellipticity}
\epsilon \approx \gamma + \epsilon_I\;,
\end{equation}
here all quantities are represented by complex numbers, $\gamma$ is the part of the ellipticity imprinted by the gravitational lensing effect of the LSS while $\epsilon_I$ denotes the intrinsic ellipticity a particular galaxy has before the light travels through the LSS. Clearly: $\langle \epsilon\rangle = 0$ by homogeneity and isotropy. Since the cosmic shear effect, $\gamma$, is at most a per cent change in shape for individual galaxies, the intrinsic ellipticity of the galaxies, $\epsilon_I$, will dominate the signal entirely. However, if galaxies are randomly oriented coherent correlations induced by $\gamma$ can be measured in the form of correlation functions or power spectra (or higher order statistics). On the other hand, if the intrinsic shapes of galaxies are correlated with the LSS there will exist correlations between the intrinsic shapes which can mimic a lensing effect, this is called $II$-alignment. Furthermore, the intrinsic shapes will also be correlated with the gravitational lensing effect itself. This latter effect is called $GI$-alignment. In this section we will review how to calculate these different contribution.
In this section we will review how to calculate these different contribution.

\subsection{Cosmic shear}
\label{sub_sec:cosmic_shear}
The lensing potential in tomographic bin $i$, $\psi_i$ \citep[for a review see e.g.][]{bartelmann_weak_2001,kilbinger_cosmology_2015} is a line-of-sight projection, i.e. a long redshift, $z$, or corresponding comoving distance $\chi(z)$, of the gravitational potential $\Phi$:
\begin{equation}\label{eq:lensing_potential}
\psi_i(\hat{\boldsymbol{n}})= \int_0^{\chi_\mathrm{H}}\mathrm{d}\chi W_i(\chi)\Phi(\hat{\boldsymbol{n}},\chi)\;,
\end{equation}
into observed direction $\hat{\boldsymbol{n}}$ which is a unit radial vector. The convergence $\kappa$ is given by a Poisson equation $\Delta \psi = 2\kappa$, where the Laplacian acts perpendicular to the line-of-sight. Observationally the lensing effect is estimated by measuring the ellipticities of the background galaxies as indicated in \cref{eq:observed_ellipticity}. In the weak lensing regime the ellipticity is an unbiased estimator of the complex shear, $\gamma$ assuming there are now intrinsic ellipticity correlations. Statistically $\kappa$ and $\gamma$ carry the same information and their angular power spectra can be shown to be equal. For simplicity we will therefore work with the convergence which. The weight function in Eq. (\ref{eq:lensing_potential}) is given by
\begin{equation}
\label{eq:weight function}
W_i(\chi) = 2 \frac{D_+(\chi)}{a}G_i(\chi)\chi\;,
\end{equation}
with the lensing efficiency function
\begin{equation}
\label{eq:lensing_efficiency}
G_i(\chi) = \int_{\mathrm{min}(\chi,\chi_i)}^{\chi_{i+1}}\mathrm{d}\chi\:n(\chi')\frac{\mathrm{d}z}{\mathrm{d}\chi'}\left(1-\frac{\chi}{\chi'}\right)\;. 
\end{equation}
with $\mathrm{d}z/\mathrm{d}\chi' = H(\chi')/c$ and $n(\chi')$ being the  distribution of the sources. Using the Limber projection \citep{limber_analysis_1954} and the Poisson equation, the angular power spectrum of the convergence in tomographic bins $i$ and $j$ can be expressed as
\begin{equation}\label{eq:limber_cosmic_shear}
C_{\kappa_i\kappa_j}(\ell) = \frac{9\Omega_\mathrm{m}^2}{4\chi_H^4}\int \frac{\mathrm{d}\chi}{\chi^2} W_i(\chi)W_j(\chi)P_\delta\left(\frac{\ell+0.5}{\chi},\chi\right)\;,
\end{equation}
with the total matter power spectrum $P_\delta (k,a(\chi))$, and the comoving Hubble radius $\chi_H$.

\subsection{Intrinsic alignments}
\label{sub_sec:intrinsic alignments}
For the intrinsic shape correlations we will assume ellipticities of galaxies to be influenced by local operators acting on the gravitational potential $\phi$. In particular we will work with the tidal-torquing model and the tidal-shearing model which apply four and two derivatives to the gravitational potential respectively. Quite generally this procedure could be generalized very similar to the bias expansion \citep{desjacques_large-scale_2018}. This has been for example done in \citep{blazek_beyond_2017}. Throughout we will assume the statistics of the density field to be well described by a Gaussian random field.

\subsubsection{Statistical properties of the large-scale structure}
We denote the gravitational tidal field as $\phi_{\alpha\beta}(\boldsymbol{x}) \equiv \partial_{\alpha}\partial_{\beta}\phi(\boldsymbol{x})$, with $\partial_\alpha \equiv \partial/\partial \boldsymbol{x}^\alpha$. The corresponding correlation function is given by:
\begin{equation}
C_{\alpha\beta\gamma\delta}(r)\equiv\bra\phi_{\alpha\beta}(\boldsymbol{x})\phi_{\gamma\delta}(\boldsymbol{x}^\prime)\ket
\end{equation}
which takes the following form \citep{catelan_correlations_2001} 
\begin{equation}
\begin{split}
C_{\alpha\beta\gamma\delta}(r)  = & \
(\delta_{\alpha\beta}\delta_{\gamma\delta}+\delta_{\alpha\gamma}\delta_{\beta\delta}+\delta_{\alpha\delta}\delta_{\beta,\gamma})\:\zeta_2(r) \\ 
& + (\hat{r}_\alpha \hat{r}_\beta \delta_{\gamma\delta}+\mathrm{5~perm.})\zeta_3(r)+
\hat{r}_\alpha \hat{r}_\beta \hat{r}_\gamma \hat{r}_\delta\:\zeta_4(r)\;,
\label{eqn_decomp}
\end{split}
\end{equation}
where $r\coloneqq |\boldsymbol{x}-\boldsymbol{x}^\prime|$ and $\hat{r}\coloneqq \boldsymbol{r}/r$.
The function $\zeta_n(r)$ encodes the statistics of $\phi$, \cite{crittenden_spin-induced_2001}:
\label{sec:quadalign}
\begin{equation}
\zeta_n(r) = \left(-1\right)^n r^{n-4}\int\frac{\de{k}}{2\pi^2}\:P_\phi(k)\,k^{n+2}\,j_n(kr)\;.
\label{eq:zeta_n}
\end{equation}
For the intrinsic shape correlation only the traceless part of the gravitational tidal field is important. Its correlation function  $\tilde{C}_{\alpha\beta\gamma\delta}(r)$ of the traceless tidal shear $\tilde{\phi}_{\alpha\beta} = \phi_{\alpha\beta} - \Delta\phi/3\:\times\delta_{\alpha\beta}$ is given by
\begin{equation}
\begin{split}
\tilde{C}_{\alpha\beta\gamma\delta}(r)&  =  \ 
C_{\alpha\beta\gamma\delta}(r) \\
& - \frac{1}{3}\left(\delta_{\gamma\delta}\left(5\zeta_2(r)+\zeta_3(r)\right) + \hat{r}_\gamma \hat{r}_\delta\left(7\zeta_3(r)+\zeta_4(r)\right)\right)\delta_{\alpha\beta} \\
& - \frac{1}{3}\left(\delta_{\alpha\beta}\left(5\zeta_2(r)+\zeta_3(r)\right) + \hat{r}_\alpha \hat{r}_\beta\left(7\zeta_3(r)+\zeta_4(r)\right)\right)\delta_{\gamma\delta} \\
& + \frac{1}{9}\left(15\zeta_2(r)+10\zeta_3(r)+\zeta_4(r)\right)\delta_{\alpha\beta}\delta_{\gamma\delta}\; .
\end{split}
\end{equation}
Furthermore, the traceless, unit-normalised gravitational tidal field $\hat{\phi}_{ij}$ can be decomposed into correlations $\tilde{C}_{AB} = \bra\tilde{\phi}_A\tilde{\phi}_B\ket$ of the traceless tidal shear $\tilde{\phi}_A$ field \citep{natarajan_angular_2001, crittenden_spin-induced_2001} by virtue of Wick's theorem,
\begin{equation}\label{eq:fourpoint}
\bra\hat{\phi}_A(\boldsymbol{x})\hat{\phi}_B(\boldsymbol{x})\:\hat{\phi}_C(\boldsymbol{x}^\prime)\hat{\phi}^\prime_D(\boldsymbol{x}^\prime)\ket =
\frac{1}{\left[14\zeta_2(0)\right]^2}\left(\tilde{C}_{AC}\tilde{C}_{BD}+\tilde{C}_{AD}\tilde{C}_{BC}\right)\;.
\end{equation}

\subsubsection{Intrinsic alignment of spiral galaxies}
In this work we will use the tidal torque model. It describes the alignment of spiral galaxies originating from their relative orientation which in turn is related to the correlation of angular momenta of neighbouring galaxies relative to the line of sight \citep{croft_weak-lensing_2000,crittenden_spin-induced_2001}. Being build up at early times, angular momentum correlations are to a large extent due to initial correlations \citep{catelan_evolution_1996, theuns_angular_1997, catelan_non-linear_1997,piras_mass_2017}. Correlated angular momenta result into correlated inclination angles and thus into correlated shapes \citep{catelan_intrinsic_2001}. By assuming the symmetry axis of the galactic disc coinciding with the direction of the angular momentum $\hat{L} = \boldsymbol{L}/L$, one can write the ellipticity as
\begin{equation}\label{eq:ellipticity_angular}
\epsilon = \frac{\hat{L}^2_x - \hat{L}^2_y}{1+ \hat{L}^2_z} + 2\mathrm{i}\frac{\hat{L}_x\hat{L}_y}{1+\hat{L}^2_z}\; .
\end{equation} 
It should be noted that this assumption is the backbone of the tidal torquing model and its applicability to weak lensing studies. In \citep{zjupa_angular_2017} and other works the spin distribution of halos was studied. The results indicate that the angular momentum of the baryonic component of a galaxy is not strongly correlated with its dark matter counterpart. This indicates that the tidal torquing model is not very well suited for the description of intrinsic alignments. In the tidal torquing theory \citep{white_angular_1984,barnes_angular_1987,schaefer_review:_2009,stewart_angular_2013}. the auto-correlations for Gaussian random fields are given by \citep{lee_galaxy_2001}
\begin{equation}
\left\langle \hat{L}_\alpha \hat{L}_\beta \right\rangle =  \frac{1}{3}\left(\frac{1+ A}{3}\delta_{\alpha\beta} - A \hat\Phi_{\alpha\mu}\hat\Phi_{\mu\beta} \right)\; .
\end{equation}
The ellipticity can finally be expressed in terms of the tidal field:
\begin{equation}
\label{eq:spiral_galaxies}
\epsilon(\hat\Phi) = 
\frac{A}{2}
\left(\hat{\Phi}_{x\alpha}\hat{\Phi}_{\alpha x} - \hat{\Phi}_{y\alpha}\hat{\Phi}_{\alpha y} -2\mathrm{i}\hat{\Phi}_{x\alpha}\hat{\Phi}_{\alpha y}\right)\; .
\end{equation}
Here the constant $A$ describes the coupling strength of the angular momentum to the shape and we assume a fiducial value of $A = 0.25$. For a more detailed discussion on this alignment model we refer to \cite{tugendhat_angular_2017,tugendhat_statistical_2018}.

\subsubsection{Intrinsic alignments of elliptical galaxies}
Elliptical galaxies can be described as  virialised system in which stars move randomly the gravitational potential $\phi$ with a characteristic velocity dispersion $\sigma^2$. In presence of tidal gravitational fields
the potential is perturbed, changing the distribution of stars and thus the observed isophotes. At first order in the gravitational tidal fields the density assumes the following form
\begin{equation}
\rho \propto \exp\left(-\frac{\phi(\boldsymbol{x})}{c^2}\right)\left(1-\frac{1}{2\sigma^2}\phi_{\alpha\beta}\big|_{\boldsymbol{x} = \boldsymbol{x_0}}x^\alpha x^\beta\right)\;.
\end{equation}
Being linear in the gravitational tidal field the model is usually referred to as the linear alignment model \citep{hirata_intrinsic_2010, blazek_beyond_2017, blazek_separating_2012, blazek_testing_2011, blazek_tidal_2015}. The coupling strength is absorbed in the constant $D$.
If $x$ and $y$ are coordinates perpendicular to the line of sight the complex ellipticity is given by
\begin{equation}
\label{eq:elliptical_galaxies}
\epsilon = 
\epsilon_+ + \epsilon_\times = 
D\left(\phi_{xx}-\phi_{yy} +2\mathrm{i}\phi_{xy}\right)\;.
\end{equation}
As a fiducial parameter we choose $D = 9.5\times10^{-5} c^2$. It should be noted that for spiral galaxies there is no definitive measurement of ellipticity correlations. However, for elliptical galaxies there is \citep{mandelbaum_detection_2006, mandelbaum_wigglez_2011}.

\subsubsection{Angular power spectra}
Angular $E$-mode ellipticity power spectra are now calculated as
\begin{equation}
\label{eq:IA_power_spectra}
C^{ij}(\ell) = \pi\int\mathrm{d}\theta \;\theta\left(C^+_{\epsilon_i\epsilon_j}(\theta)J_0(\ell\theta)+C^-_{\epsilon_i\epsilon_j}(\theta)J_4(\ell\theta)  \right)\;,
\end{equation}
where $J_i$ are Bessel functions and $C^\pm_{\epsilon_i\epsilon_j}$ is defined as
\begin{equation}
C^\pm_{\epsilon_i\epsilon_j} \coloneqq \langle \epsilon_{+,i}\epsilon^\prime_{+,j}\rangle \pm \langle \epsilon_{\times,i}\epsilon^\prime_{\times,j}\rangle\;.
\end{equation}
Here we split the complex ellipticity $\epsilon = \epsilon_+ + \epsilon_\times$. Note that $\epsilon_i$ is a placeholder for either the intrinsic ellipticity component of spiral or elliptical galaxies as defined in Eq. (\ref{eq:spiral_galaxies}) and Eq. (\ref{eq:elliptical_galaxies}) respectively, or for the lensing signal. What follows are the three contributions on top of the lensing signal: two $II$ contributions from both galaxy types and a single $GI$ contribution from elliptical galaxies.
\begin{figure}
\begin{center}
\includegraphics[width=0.99\textwidth]{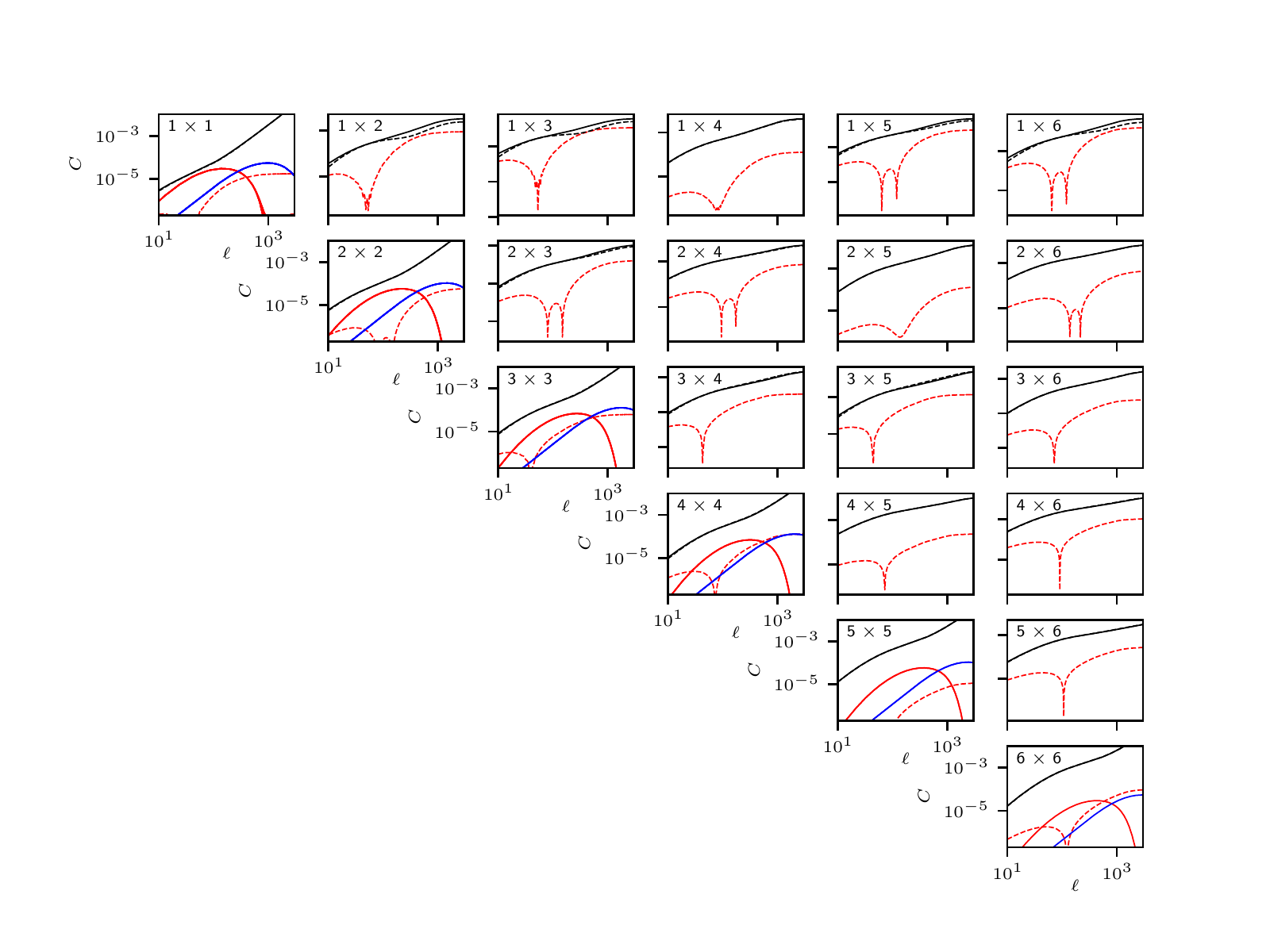}
\caption{Components of the observed intrinsic alignment angular power spectra in all tomographic bins. The solid black line shows the lensing signal including the shape noise contribution. Solid red and blue lines show the $II$ contribution from ellipticals and spirals respectively. The dashed red line shows the $GI$ contribution from ellipticals and the dashed black line is the sum of all contributions. It should be noted that the $GI$ contribution experiences a sign flip.}
\label{fig:corner_spectra}
\end{center}
\end{figure}
For the redshift distribution $n(z)\dd z$ is assumed to have the shape,
\begin{equation}
n(z)\dd z \propto \left(\frac{z}{z_0}\right)^2\exp\left[-\left(\frac{z}{z_0}\right)^\beta\right]\dd z,
\end{equation}
with the choices $\beta=3/2$ and $z_0 = 0.9$, which generates a median redshift of unity \citep{laureijs_euclid_2011}. Furthermore, we assume $n_\mathrm{gal}=30\;\mathrm{arcmin}^{-2}$ and a sky fraction of $f_\mathrm{sky} = 0.3$.
For completeness we show the corresponding spectra in \autoref{fig:corner_spectra} for six tomographic bins and refer to \citep{tugendhat_angular_2017} for further details.

\section{Statistics}
\label{sec:stat}
\subsection{Likelihood}
In our forecast we assume the modes of the convergence field to follow a Gaussian distribution with zero mean
\begin{equation}\label{eq:gaussian}
\boldsymbol{\kappa}_{\ell m} \sim \mathcal{N}(\boldsymbol{0},\boldsymbol{C}_{\ell})\;,
\end{equation}
where the components of the covariance matrix $\left(\boldsymbol{C}_\ell\right)_{ij}$ are given by the tomographic ellipticity spectra. The data-averaged logarithmic likelihood is in this case given by
\begin{equation}\label{eq:chi2}
L (\boldsymbol{\theta}) \coloneqq -2\log \mathcal{L}= f_\mathrm{sky}\sum_{\ell=\ell_\mathrm{min}}^{\ell_\mathrm{max}}\frac{2\ell+1}{2} \left[ \log\left(\frac{\det\boldsymbol{C}}{\det\boldsymbol{\hat{C}}}\right) + \mathrm{tr}\left(\boldsymbol{C}^{-1}\boldsymbol{\hat{C}}\right) -n_\mathrm{tomo}\right]\;,
\end{equation}
where we made the dependence on the cosmology, $\boldsymbol{\theta}$ explicit and arbitrarily shifted $L$ such that $L(\boldsymbol{\theta})|_{\boldsymbol{\theta}_0}=0$, where ${\boldsymbol{\theta}_0}$ is the fiducial cosmology. $\boldsymbol{\hat{C}}$ is the covariance matrix at the fiducial cosmology.
Finally, observed angular power spectra include a shape noise contribution due to the finite amount of source galaxies. Thus:
\begin{equation}
C_{\kappa_i\kappa_j}\to C_{\kappa_i\kappa_j} + \frac{\sigma_\epsilon ^2n_\mathrm{tomo}}{\bar n}\delta_{ij}\;,
\end{equation}
where $\bar n$ is the average number of galaxies per steradian and $\sigma_\epsilon ^2\approx 0.3$ is the intrinsic ellipticity dispersion of galaxies.
By virtue of Bayes' theorem, for which we assume a flat prior if not otherwise specified, we sample from (\ref{eq:chi2}) using affine invariant sampling techniques \cite{goodman_ensemble_2010}. 

We set up the following situation: The observed data, i.e. the covariance matrix at the fiducial cosmology $\boldsymbol{\hat{C}}$ has the following contributions
\begin{equation}
\label{eq:fiducial_covariance}
(\boldsymbol{\hat{C}})_{ij} = C_{\kappa_i\kappa_j} + C^{\epsilon_i\epsilon_i}_{\mathrm{spiral}}\delta_{ij}+ C^{\epsilon_i\epsilon_i}_{\mathrm{elliptical}}\delta_{ij} + C^{\epsilon_i\kappa_j}_{\mathrm{elliptical}}\;.
\end{equation}
That is, we set up the true model including intrinsic alignments for spiral and elliptical galaxies. This is done such that we assume $70$ per cent of the galaxy sample to be spirals and the remaining to be ellipticals. For more details on this we refer to \citep{tugendhat_angular_2017}. We then fit to this fiducial cosmology a cosmology which only contains the lensing signal, i.e. $(\boldsymbol{C})_{ij} =C_{\kappa_i\kappa_j}$. 

\begin{center}
    \begin{figure}
        \centering
        \includegraphics{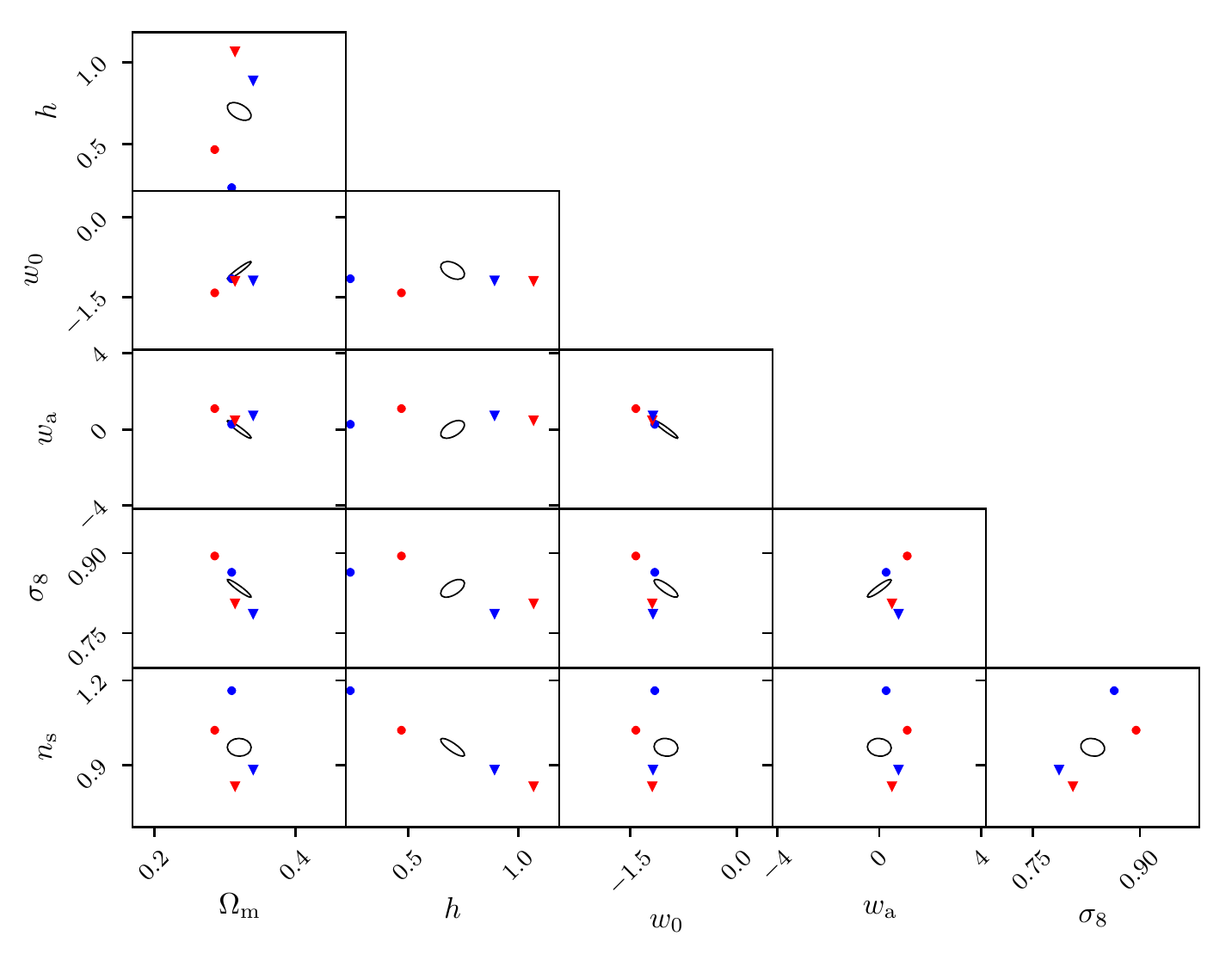}
        \caption{Fisher forecast for a \texttt{Euclid}-like survey, black ellipses correspond to the $1\sigma$ contour. The circles mark the systematic bias estimated by \cref{eq:gaussian_bias} using \cref{eq:G_and_a_for_gaussian}, while the triangles correspond to he bias as estimated using the wrong equation (\cref{eq:G_old}, cf \citep{tugendhat_angular_2017}). Red corresponds to GI and II alignment, while blue only assumes II alignment.}
        \label{fig:Fisher_bias}
    \end{figure}
\end{center}

\subsection{Parameter estimation bias}
We will reconsider the calculation described in \citep{schaefer_weak_2012} and use Gaussian likelihoods where the parameter dependence is carried by the covariance and $\langle\bs{x}\rangle = 0$, where the brackets denote an ensemble average. The general form of the Gaussian logarithmic likelihood of a data vector $\bs{x}$ and a parameter vector $\bs{y}$ is given by
\begin{equation}
\label{eq:Gaussian_likelihood}
\mathcal{L} \equiv -2\log[p(\bs{x}|\bs{y})] = \mathrm{tr}\left[\log\bs{C} + \bs{C}^{-1}\bs{D}\right]\;,
\end{equation}
where $\bs{D} = \bs{x}\otimes\bs{x}$ is the data covariance and $\bs{C}$ is the covariance matrix. 
Following \citep{schaefer_weak_2012} we consider $\mathcal{L}_t$ and $\mathcal{L}_f$, i.e. the true and the wrong logarithmic likelihood respectively. Furthermore, we will use the following shorthand notation:
\begin{equation}
\label{eq:notation}
\left(\frac{\partial f}{\partial \bs{y}}\right)_\mu \coloneqq \frac{\partial f}{\partial y^\mu} \equiv f_{,\mu}\;. 
\end{equation}
With this we note that
\begin{equation}\label{eq:extremum_constraint}
\left\langle\mathcal{L}_{t,\mu}\big|_{\bs{y}_t}\right\rangle = 0 = \left\langle\mathcal{L}_{f,\mu}\big|_{\bs{y}_f}\right\rangle\;,
\end{equation}
due to the maximum constraint, where $\bs{y}_t$ and $\bs{y}_f$ are the best fit values of the true and the wrong likelihood respectively. Here we use the shorthand notation $f(x)|_{x=x_0} \equiv f|_{x_0}$.
Now, we expand $\mathcal{L}_f$ around $\bs{y_t}$, i.e.
\begin{equation}\label{eq:likelihood_expansion}
\mathcal{L}_f(\bs{y}) \approx \mathcal{L}_f\big|_{\bs{y}_t} + \mathcal{L}_{f,\mu}\big|_{\bs{y}_t}(\bs{y} - \bs{y}_t)^\mu + \frac{1}{2}\mathcal{L}_{f,\mu\nu}\big|_{\bs{y}_t} (\bs{y} - \bs{y}_t)^\mu (\bs{y} - \bs{y}_t)^\nu + \dots \; ,
\end{equation}
where the sum convention is implied.
One can now make use of Eq. (\ref{eq:extremum_constraint}) to find
\begin{equation}\label{eq:expansion_bias}
\left\langle\mathcal{L}_{f,\mu}(\bs{y})\big|_{\bs{y}_f}\right\rangle = 0 = \left\langle\mathcal{L}_{f,\mu}\big|_{\bs{y}_t}\right\rangle + \left\langle  \mathcal{L}_{f,\mu\alpha}\big|_{\bs{y}_t}\delta^\alpha \right\rangle + \frac{1}{2}\left\langle \mathcal{L}_{f,\mu\alpha\beta}\big|_{\bs{y}_t} \delta^\alpha\delta^\beta \right\rangle +\dots\; .
\end{equation}
Here we defined the bias $\bs{\delta} \coloneqq \bs{y}_f-\bs{y}_t$. If the logarithmic likelihood is at most quadratic in the parameters, as it is the case for a Gaussian posterior distribution, Eq. (\ref{eq:expansion_bias}) can be cast into a linear system of equations with the solution
\begin{equation}\label{eq:gaussian_bias}
\bs{\delta} = \bs{G}^{-1} \bs{a}, \quad \mathrm{with} \quad G_{\mu\alpha} = -\left\langle  \mathcal{L}_{f,\mu\alpha}\big|_{\bs{y}_t}\right\rangle 
 \quad \mathrm{and} \quad a_\mu =\left\langle\mathcal{L}_{f,\mu}\big|_{\bs{y}_t}\right\rangle\;
\end{equation}
For the likelihood (\ref{eq:Gaussian_likelihood}) one finds the following expressions
\begin{equation}
\label{eq:G_and_a_for_gaussian}
\begin{split}
a_\mu = & \ \mathrm{tr}\left[\bs{C}^{-1}\bs{C}_{,\mu}(\bs{\mathrm{id}} -\bs{C}^{-1}\bs{C}_t)\right]\;,\\
G_{\mu\nu} = &  \ \mathrm{tr} \left[\bs{C}^{-1}\bs{C}_{,\mu\nu} (\bs{C}^{-1}\bs{C}_t-\bs{\mathrm{id}}) +  \bs{C}^{-1}\bs{C}_{,\mu}\bs{C}^{-1}\bs{C}_{,\nu}(\bs{\mathrm{id}} - \bs{C}^{-1}\bs{C}_t) -   \bs{C}^{-1}\bs{C}_{,\nu}\bs{C}^{-1}\bs{C}_{,\mu} \bs{C}^{-1}\bs{C}_t)  \right]\;,
\end{split}
\end{equation}
where $\bs{C}$ is to be evaluated at $\bs{y}_t$ and we used that $\langle\bs{D}\rangle = \bs{C}_t$. Lastly, $\bs{\mathrm{id}}$ is the identity. It should be noted that if $\bs{y}_t = \bs{y}_f$, $-G_{\mu\nu}$ reduces to the usual Fisher matrix. Furthermore, this result differs slightly from other results in the literature, since one cannot commute the derivative with respect to the $\mu$-th and $\nu$-th component in the last two terms.

\section{Results}
\label{sec:results}
\subsection{Analytic parameter estimation bias}
Initially, we repeat a similar analysis as was shown in \citep{tugendhat_angular_2017}. However, it should be noted that we include an additional parameter in the analysis, $w_\mathrm{a}$, as well as six instead of five tomographic bins.
In particular we investigate the parameter estimation bias using the analytical formulas derived in section \ref{sec:stat} (see also \citep{schaefer_weak_2012}). Taking into account a sum over all multipoles $\ell$, as well as the additional multiplicity factor from the $m$ modes, $(2\ell+1)/2$, we calculate the bias using \cref{eq:G_and_a_for_gaussian}. Additonally we use another expression for $\boldsymbol{G}$ \citep{schaefer_weak_2012,tugendhat_angular_2017}:
\begin{equation}
\label{eq:G_old}
G_{\mu\nu} =   \ \mathrm{tr} \left[\bs{C}^{-1}\bs{C}_{,\mu\nu} (\bs{C}^{-1}\bs{C}_t-\bs{\mathrm{id}}) +  \bs{C}^{-1}\bs{C}_{,\mu}\bs{C}^{-1}\bs{C}_{,\nu}(\bs{\mathrm{id}} - 2\bs{C}^{-1}\bs{C}_t)  \right]\;,
\end{equation}
as discussed in the previous section. \Cref{fig:Fisher_bias} shows the biases together with the Fisher matrix forecast contours for an unbiased measurement. Coloured circles represent the analytic bias using \cref{eq:G_and_a_for_gaussian}, while coloured triangles show the bias obtained from \cref{eq:G_old}. Finally, red indicates that the true covariance $\boldsymbol{C}_t$ is biased with both II and GI alignment terms, while blue shows the case where only the II alignment terms give rise to a bias. Clearly, both expressions do not agree as expected. The calculated biases differ substantially and even point in the opposite direction for a few parameters. For almost all parameters the bias is much bigger than the average statistical error, raising the question of the applicability of the approximation done in \cref{eq:likelihood_expansion}. Generally it is not clear how accurate the approximation is in cases where $\sum_{\mu\nu}F_{\mu\nu}\delta^\mu\delta^\nu \gg 1$. In \Cref{app:ngbias} we discuss a possible extension of the formalism to higher order using a derivative expansion. However, we find only marginal improvement over the Gaussian results due to the large systematics in the spectra involved. It is thus necessary to invoke a direct sampling of the full posterior using MCMC.

\begin{center}
    \begin{figure}
        \centering
        \includegraphics{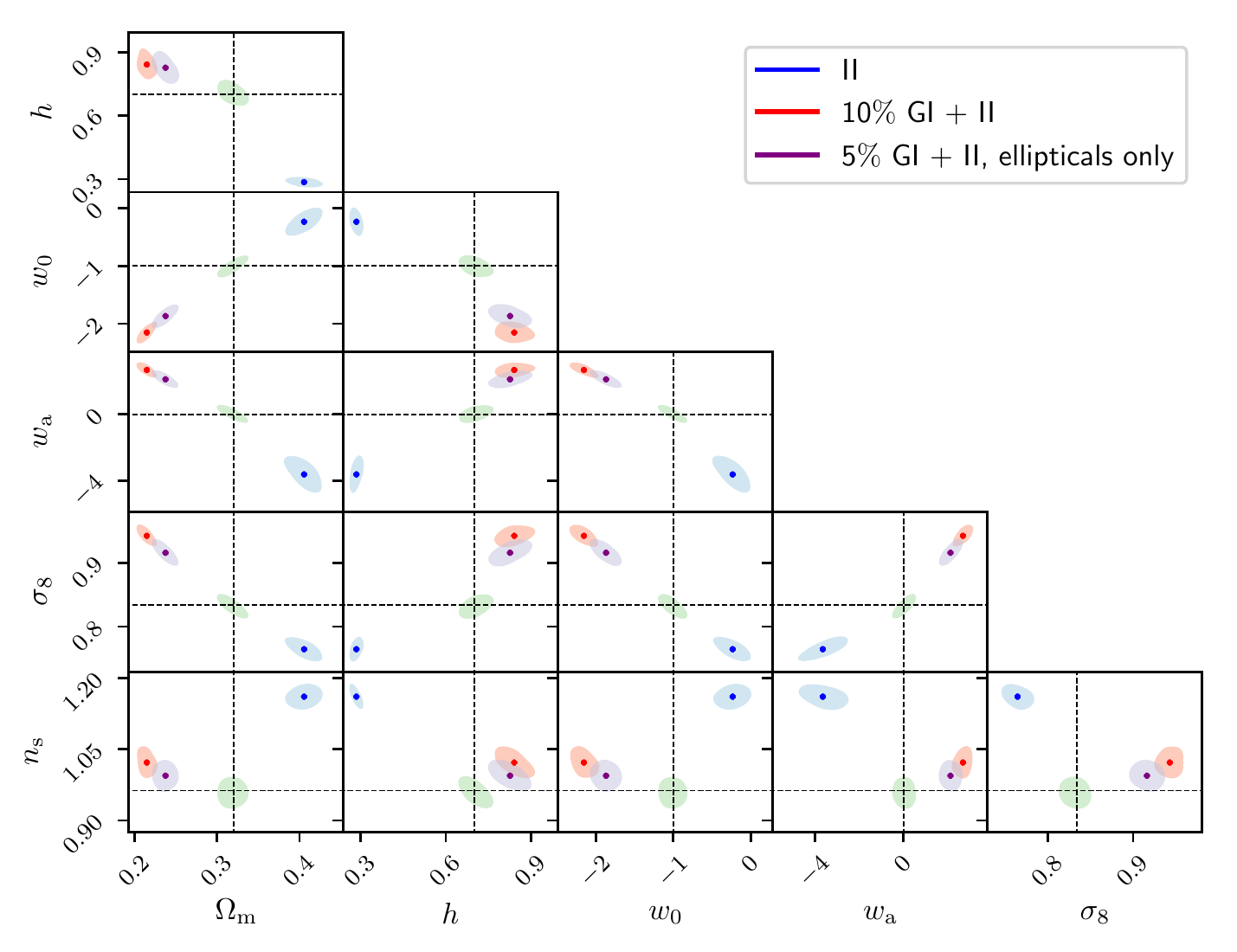}
        \caption{$1\sigma$ contours as estimated from MCMCs. The green contours shows the constraints at the fiducial cosmology without any intrinsic alignment present. Blue contours correspond to II alignment only. Purple contours show the contours when five per cent residual intrinsic alignments by elliptical galaxies are present in the survey. Lastly, red contours show the case where both GI and II alignment are present. However, in the latter case the strength of the alignment was reduced by $90\%$ for both spirals and ellipticals. The dashed lines mark the fiducial cosmology.}
        \label{fig:II_alignment_bias}
    \end{figure}
\end{center}

\subsection{Parameter estimation bias from MCMC}
Here we discuss the impact of II and GI alignments on parameter estimation bias which is obtained directly from MCMC methods. We sample from the likelihood using affine invariant sampling suggested in \citep{goodman_ensemble_2010}.

To begin with, we show the influence of II and GI alignment of fixed amplitude on the parameter inference process. In \cref{fig:II_alignment_bias} contours for a \texttt{Euclid}-like survey are shown in green. The blue contours depict a weak lensing measurement which is still plagued by an II alignment contribution. Comparing this to \cref{fig:Fisher_bias} one can see clear differences in the predicted bias, in particular the bias is underestimated for almost all parameters, showing that the approximations used to calculate the analytical bias is not suited any more. Interestingly for certain parameters the bias actually switches sign. In red we show a measurement where $\boldsymbol{C}_t$ contains GI and II alignment, however with its amplitude reduced by 90 per cent. Clearly, even for this strongly reduced alignment signal, the parameter estimation bias can be as large as calculated by the analytic expression when using the 100 per cent of the intrinsic alignment signal. Therefore, we conclude that the analytic expression is not suitable to estimate the parameter estimation bias in this case. The reason for this is that the intrinsic alignment signal introduces a different $\ell$ and tomographic bin dependence than the pure lensing signal. If these dependences would be very similar, the analytic expression would yield much better results since it would only need to account for a changing amplitude in the covariance. For the GI-terms the situation becomes even worse, since GI induces an anti-correlation and therefore reduces the observed signal in the cross-correlation of different bins. This situation is very difficult to fit for the six parametric model we used here and wide regions of the parameter space have to be explored, leading to the Gaussian approximation of the posterior being insufficient. 
In purple we show the a situation where only intrinsic alignments by elliptical galaxies are present with the amplitude reduced by 95 per cent. Clearly, the systematic biases are completely dominated by the GI alignment which, unfortunately, is notoriously difficult to remove from surveys since it is a non-local effect. Furthermore, we see that the II and GI terms can give rise to biases with opposite sign, thus partially cancelling each other.

\begin{center}
    \begin{figure}
        \centering
        \includegraphics[width = 0.47\textwidth]{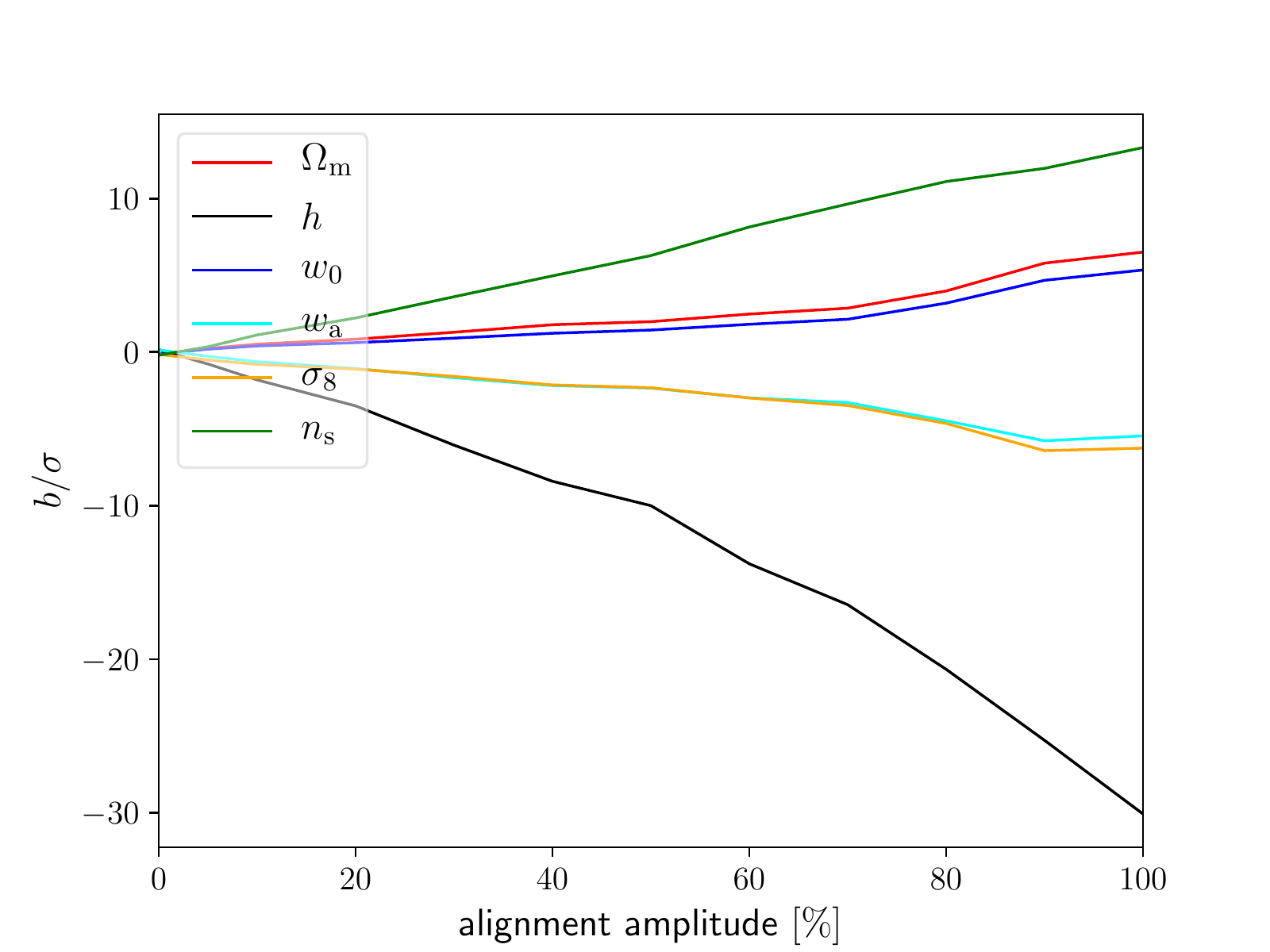}
        \includegraphics[width = 0.47\textwidth]{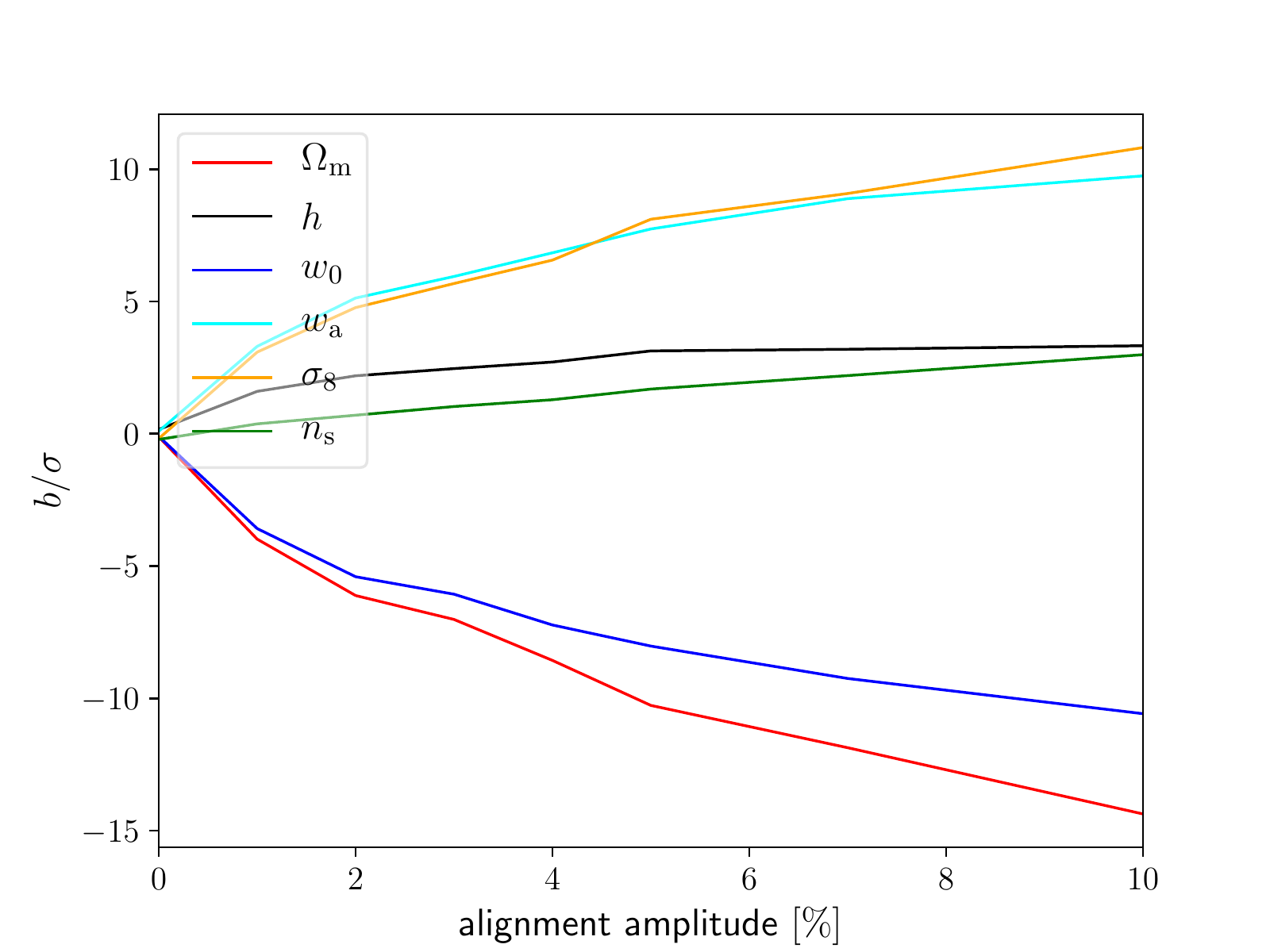}
        \caption{Relative bias, i.e. the bias $b$ scaled by the marginalized $1\sigma$ error, as a function of the alignment amplitude in per cent relative to its fiducial value. Different lines correspond to the different fitted cosmological parameters. The left column shows again the case where only II alignment is present, while the right plot includes both II and GI alignment. Note the different axis range.}
        \label{fig:MCMC_relative_bias}
    \end{figure}
\end{center}

\Cref{fig:MCMC_relative_bias} shows the relative bias, $b/\sigma$ for the six cosmological parameters used in this work. The left figure represents the influence of the II terms. The parameters biased the strongest are $h$ and $n_\mathrm{s}$. This can be understood from the fact that the II alignment mimics a stronger lensing signal which is strongest in the tomographic bins at low redshift. In turn, the model tries to fit this additional power by increasing $n_\mathrm{s}$, which can partially account for the redshift dependence due to the projection along the line of sight. At the same time the additional amplitude can effectively be reduced by reducing $h$. As soon as GI alignments are present as well, the situation changes dramatically, as shown in the right panel of \cref{fig:MCMC_relative_bias}. In this more realistic case $\Omega_\mathrm{m}$ alongside with the two dark energy equation of state parameters $w_0$ and $w_\mathrm{a}$ are showing the largest bias. As explained before the model now has to reduce the signal in the cross-correlations, especially at high $\ell$, but at the the time take into account the enhanced signal for auto-correlations. In particular it is very illustrative that both $w_0$ and $w_\mathrm{a}$ can be biased by roughly $5\sigma$ even if only one per cent of the initial alignment signal remains in the data. 

Finally, in our analysis we always used the non-linear power spectrum as predicted by \texttt{Halofit} and summed up all scales theoretically accessible to \texttt{Euclid}. There are three things to question here: $(i)$ How accurate is the prediction of the matter power spectrum at these small scales; $(ii)$ how well does the \texttt{Halofit} prescription describe the change of the matter power spectrum with the cosmological parameters considered, especially in the regions of parameter space explored here; $(iii)$ How well is the intrinsic shape correlation of both galaxy types described at small scales. 
The first one, for our purpose, is the least important, since we can just assume that for a given fiducial cosmology, the power spectrum actually describes the real world well enough. In order to address $(ii)$ and $(iii)$ we carry out an analysis in \Cref{app:bias_linear} where we remove all non-linear scales from the survey, thus ignoring the problematic parts of the power spectrum modelling. Furthermore, on linear scales the dominant alignment contribution arises from elliptical galaxies, which have been studied extensively in simulations \citep{zjupa_intrinsic_2019}. For this setting we still find strong systematic biases in all parameters roughly half as large as for the full multipole range.

\section{Conclusion and Discussion}
\label{sec:conclusions}
In this paper we have investigated the impact of IA on the parameter inference process with a weak gravitational lensing signal such as \texttt{Euclid}. For our analysis we assumed a Gaussian likelihood for the convergence modes. The cosmological model used was a standard six parameter $w$CDM model as typical for weak lensing surveys. Matter power spectra have been calculated with \texttt{Halofit} and the Limber approximation was employed. For the IA signal we assumed tidal torquing and shearing for spiral and elliptical galaxies respectively. The most critical assumption here is the tidal torquing model, which equates the spin of dark matter halos with the one of the visible matter. This relationship might be very complicated and depends strongly on the halo formation history. However even if the alignment signal by spiral galaxies is much weaker than modelled here our main results do not change significantly as the largest contamination arises from elliptical galxies whose intrinsic alignment model is much more roboust. Finally we sampled from the Gaussian sampling distribution using MCMC analysis calculate the parameter estimation bias and investigated its dependence on the residual alignment signal in the analysis. 
Our main results can be summarized as follows:

$(i)$ The analytic formula for the parameter estimation bias \citep{schaefer_galactic_2012} used for example in \citep{tugendhat_angular_2017} is a symmetrized version of the correct equation, which does not allow for the identification of two terms, reflecting the non-symmetry of divergencies between two probability distributions. The biases derived with both equations can differ.

$(ii)$ In general, the analytic prescription is only valid for small biases and in a region of the parameter space where the Gaussian approximation is still valid, in particular it should not exceed the statistical uncertainty of the experiment. If the bias is large and the model has to explore regions of the parameter space far away from the point of the Taylor expansion, the analytic expressions breaks down. In particular it is not well suited for the biases encountered due to IA.

$(iii)$ Remaining IA signal can bias the inferred parameters in a weak lensing analysis dramatically. If both GI and II alignment, described by the models mentioned above, are present in the survey a one per cent residual alignment signal suffices to bias key parameters for lensing surveys such as the dark energy equation of state by $5\sigma$. This strong bias is mainly driven by the GI part of the alignment, which is very difficult to remove due to its non-local nature.

$(iv)$ With alignments of spiral galaxies being not detected yet one might object whether these galaxies align intrinsically at all. However, even if only elliptical galaxies align, the parameter estimation bias is of the same order of magnitude. Moreover, the presence of II alignments for spiral galaxies slightly alleviates the resulting bias within our model.

We therefore conclude that IA has to removed or modelled to well below a per cent accuracy in order to achieve the promised science goals of \texttt{Euclid}-like surveys. This also holds true if only cross-correlations between different tomographic bins are considered, since the GI contributions are particularly difficult to fit by standard parameters since they decrease the overall signal.
The results are also robust against the modelling of spiral galaxies, which itself requires more detailed studies.

\acknowledgments 
R.R. acknowledges funding by the Israel Science Foundation (grant no. 1395/16 and 255/18).

\appendix
\section{Non-Gaussian systematic bias}
\label{app:ngbias}
In order to include non-Gaussian features in the posterior distribution. One example would be to include the third term in Eq. (\ref{eq:expansion_bias}). This would amount to third derivatives of the covariance and to a quadratic equation in the bias. However, as pointed out in \citep{sellentin_fast_2015} a particularly useful expansion should not be done on the level of the logarithmic likelihood, but on the level of the spectra. We will therefore also write the covariance matrix as
\begin{equation}\label{eq:dali_expansion}
\bs{C} = \bs{C}_0 + \bs{\mathcal{T}}_C\;,
\end{equation}
where $\bs{C}_0$ the covariance at a reference point, $\bs{y}_0$, and $\bs{\mathcal{T}}_C$ its Taylor expansion starting from the first order term around that reference point:
\begin{equation}\label{eq:taylor_expansion_covariance}
\bs{\mathcal{T}}_C \coloneqq \sum_{n=1}^\infty\frac{\bs{C}_{,\{\alpha\}_n}}{n!}(\bs{y}-\bs{y}_0)^{\{\alpha\}_n}...(\bs{y}-\bs{y}_0)^{\{\alpha\}_n}\;,
\end{equation}
with $\{\alpha\}_n$ the set of all indices corresponding to the $n$th derivative. In \citep{sellentin_fast_2015} the logarithmic likelihood is then expanded to second order in $ \bs{\mathcal{T}}_C$, yielding
\begin{equation}
\label{eq:DALI_expansion}
\begin{split}
\mathcal{L}(\bs{y})  \approx& \ \mathrm{tr}\bigg[\log\bs{C} +\bs{C}^{-1}\bs{\mathcal{T}}_C  - \frac{1}{2}\bs{C}^{-1}\bs{\mathcal{T}}_C\bs{C}^{-1}\bs{\mathcal{T}}_C \\ & +  \bs{D}\left(\bs{C}^{-1}-\bs{C}^{-1}\bs{\mathcal{T}}_C\bs{C}^{-1}+\bs{C}^{-1}\bs{\mathcal{T}}_C\bs{C}^{-1}\bs{\mathcal{T}}_C\bs{C}^{-1}\right) \bigg]\;.
\end{split}
\end{equation}
We can now use this expansion instead of (\ref{eq:likelihood_expansion}) and use the wrong covariance evaluated at the true model as the reference $\bs{C}_0$. Using again the constraint (\ref{eq:extremum_constraint}) one finds, with $\bs{C}\equiv \bs{C}_f(\bs{y}_t)$ understood, the expression
\begin{equation}
\begin{split}
\label{eq:DALI_applied_constraint}
0 = \ \mathrm{tr}\bigg[ & \bs{C}^{-1}\bs{C}_{,\mu} - \bs{C}^{-1}\bs{C}_{,\mu}\bs{C}^{-1}\bs{\mathcal{T}}_C + \bs{C}^{-1}\bs{\mathcal{T}}_{C,\mu} + \bs{C}^{-1}\bs{C}_{,\mu}\bs{C}^{-1}\bs{\mathcal{T}}_C\bs{C}^{-1}\bs{\mathcal{T}}_C - \bs{C}^{-1}\bs{\mathcal{T}}_{C,\mu}\bs{C}^{-1}\bs{\mathcal{T}}_C  \\
& + \bs{C}_t\bigg( \bs{C}^{-1}\bs{C}_{,\mu}\bs{C}^{-1}\bs{\mathcal{T}}_C\bs{C}^{-1} -\bs{C}^{-1}\bs{C}_{,\mu}\bs{C}^{-1} - \bs{C}^{-1} \bs{\mathcal{T}}_{C,\mu}  \bs{C}^{-1} +  \bs{C}^{-1}\bs{\mathcal{T}}_C\bs{C}^{-1}\bs{C}_{,\mu}\bs{C}^{-1} \\
& -\bs{C}^{-1}\bs{\mathcal{T}}_C\bs{C}^{-1}\bs{C}_{,\mu}\bs{C}^{-1}\bs{\mathcal{T}}_C\bs{C}^{-1}
-\bs{C}^{-1}\bs{C}_{,\mu}\bs{C  }^{-1}\bs{\mathcal{T}}_C\bs{C}^{-1}\bs{\mathcal{T}}_C\bs{C}^{-1} \\
& -\bs{C}^{-1}\bs{\mathcal{T}}_C\bs{C}^{-1}\bs{\mathcal{T}}_C\bs{C}^{-1}\bs{C}_{,\mu}\bs{C}^{-1} 
 + \bs{C}^{-1}\bs{\mathcal{T}}_{C,\mu}\bs{C}^{-1}\bs{\mathcal{T}}_{C}\bs{C}^{-1} + \bs{C}^{-1}\bs{\mathcal{T}}_{C}\bs{C}^{-1}\bs{\mathcal{T}}_{C,\mu}\bs{C}^{-1}\;.
\bigg)\bigg] \\
=  \ \mathrm{tr}\bigg[ & \bs{C}^{-1}\bs{C}_{,\mu}\left(\bs{\mathrm{id}} -  \bs{C}^{-1}\bs{\mathcal{T}}_C + \bs{C}^{-1}\bs{\mathcal{T}}_C\bs{C}^{-1}\bs{\mathcal{T}}_C \right) +
\bs{C}^{-1}\bs{\mathcal{T}}_{C,\mu}\left(\bs{\mathrm{id}} - \bs{C}^{-1}\bs{\mathcal{T}}_C\right) \\
& +\bs{C}^{-1}\bs{C}_t\bs{C}^{-1}\bigg(\bs{C}_{,\mu}\left(\bs{C}^{-1}\bs{\mathcal{T}}_C-\bs{\mathrm{id}} - \bs{C}^{-1}\bs{\mathcal{T}}_C\bs{C}^{-1}\bs{\mathcal{T}}_C\right) +\bs{\mathcal{T}}_{C,\mu}\left(\bs{C}^{-1}\bs{\mathcal{T}}_C-\bs{\mathrm{id}}\right) \\
& + \bs{\mathcal{T}}_C\bs{C}^{-1}\left(\bs{C}_{,\mu} -\bs{C}_{,\mu} \bs{C}^{-1}\bs{\mathcal{T}}_C + \bs{\mathcal{T}}_{C,\mu} - \bs{\mathcal{T}}_C\bs{C}^{-1} \bs{C}_{,\mu}\right) 
\bigg) 
\bigg]
\end{split}
\end{equation}
Due to $\bs{C}_t \neq \bs{C}$ this expression cannot be simplified further. We now insert the Taylor expansion and consider terms up to second order only. In this case we find
\begin{equation}\label{eq:taylor_expansion_linear}
\bs{\mathcal{T}}_{C} = \bs{C}_{,\alpha}\delta^\alpha + \frac{1}{2} \bs{C}_{,\alpha\beta}\delta^\alpha\delta^\beta\;,\quad \bs{\mathcal{T}}_{C,\mu} = 2\bs{C}_{,\mu\alpha}\delta^\alpha + \bs{C}_{,\mu} +\bs{C}_{,\mu\alpha\beta}\delta^\alpha\delta^\beta;.
\end{equation}
Inserting this into (\ref{eq:DALI_applied_constraint}) we collect terms of zeroth, first and second order in the bias $\bs{\delta}$ such that
\begin{equation}
\label{eq:splitting}
0 = \mathcal{A}_\mu + \mathcal{B}_{\mu\alpha} \delta^\alpha+ \mathcal{C}_{\mu\alpha\beta} \delta^\alpha\delta^\beta\;.  
\end{equation}
In particular we find $\mathcal{A}_\mu = a_\mu$, $\mathcal{B}_{\mu\alpha} = -G_{\mu\alpha}$ and
\begin{equation}
\label{eq:components_dali_bias}
\begin{split}
\mathcal{C}_{\mu\alpha\beta} =  \ \frac{1}{2} \mathrm{tr}\bigg[&
\bs{C}^{-1}\bs{C}_{,\mu}\bs{C}^{-1}\bs{C}_{,\alpha}\bs{C}^{-1}\bs{C}_{,\beta}(\bs{\mathrm{id}} 
- \bs{C}^{-1}\bs{C}_t) + {C}^{-1}\bs{C}_t{C}^{-1}\bs{C}_{\alpha} {C}^{-1}\left(C_{,\mu} {C}^{-1}\bs{C}_{,\beta} - C_{,\beta}{C}^{-1}\bs{C}_{,\mu}\right) \\
& -2\bs{C}^{-1}\bs{C}_{,\mu\alpha}\bs{C}^{-1}\bs{C}_{,\beta}\left(\bs{\mathrm{id}} - \bs{C}^{-1}\bs{C}_t\right) -\bs{C}^{-1}\bs{C}_{,\mu}\bs{C}^{-1}\bs{C}_{,\alpha\beta}\left(\bs{\mathrm{id}} - \bs{C}^{-1}\bs{C}_t\right) \\
& + 2\bs{C}^{-1}\bs{C}_t\bs{C}^{-1}\bs{C}_{,\alpha}\bs{C}^{-1}\bs{C}_{,\beta\mu} + \bs{C}^{-1}\bs{C}_{,\alpha\beta\mu}\left(\bs{\mathrm{id}} - \bs{C}^{-1}\bs{C}_t\right)
\bigg]\,.
\end{split}
\end{equation}
This approach now obviously yields two solutions for $\delta^\alpha$, from which one the correct one can be identified by demanding that the bias vanishes for vanishing systematic. We applied this approach to the same system as in \cref{fig:Fisher_bias} but found only marginal improvement over the Gaussian case. The reason is again that the systematic drives the expanded likelihood into regions far away from its maximum where the two posteriors (the correct and the incorrect one) are compared, rendering the approach inapplicable.

\begin{center}
    \begin{figure}
        \centering
        \includegraphics[width = 0.9\textwidth]{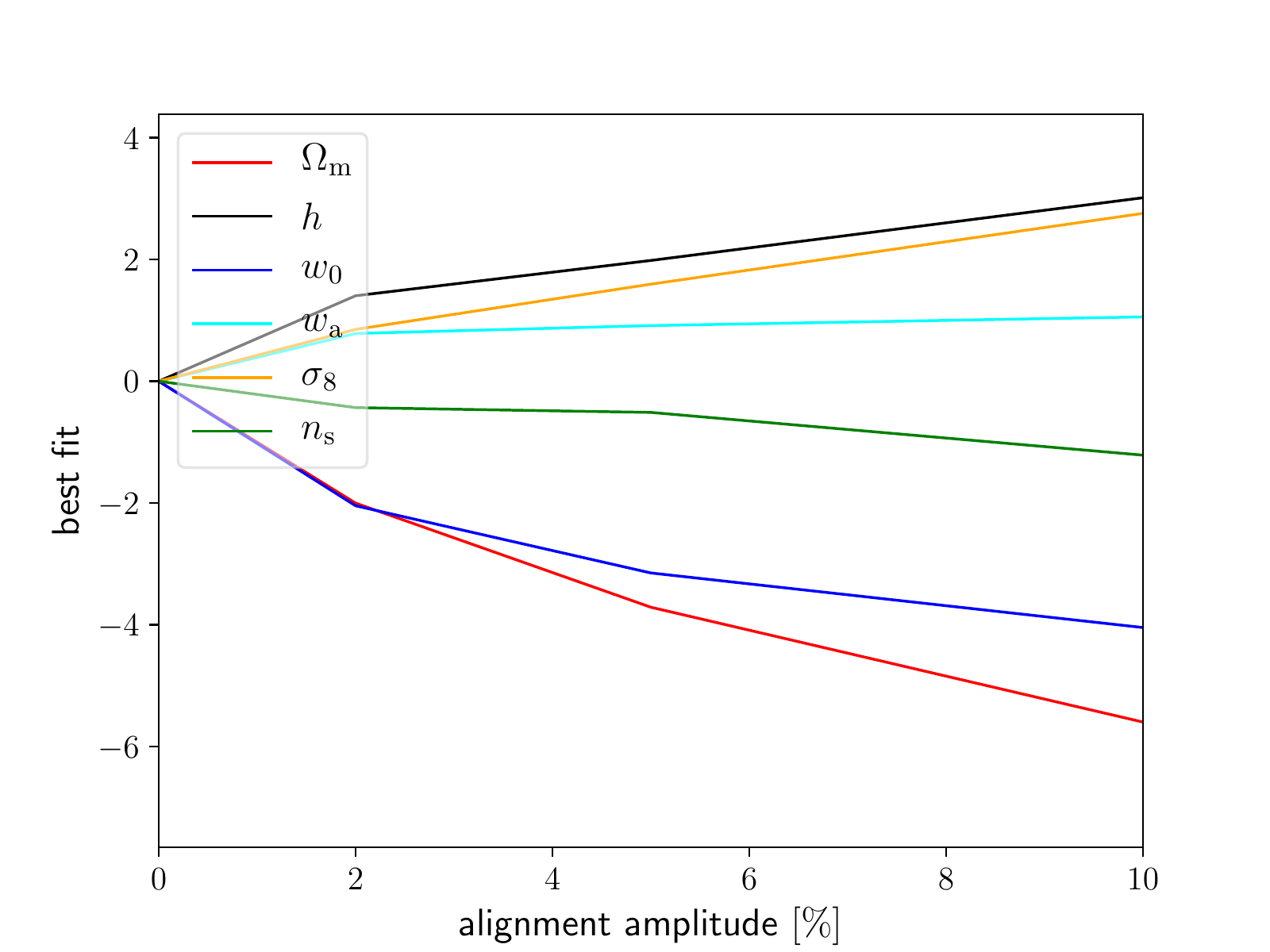}
        \caption{Relative bias, i.e. the bias $b$ scaled by the marginalized $1\sigma$ error, as a function of the alignment amplitude in per cent relative to its fiducial value. The colour scheme is the same as in \cref{fig:MCMC_relative_bias}.  Here the non-linear scales have been removed from the survey (see the discussion in \cref{sec:results}). The right panel shows the systematic bias as a function of maximum multipole $\ell_\mathrm{max}$ for a survey with 10 per cent residual alignment from both elliptical and spiral galaxies.}
        \label{fig:MCMC_relative_bias_linear_ell}
    \end{figure}
\end{center}

\section{Bias from linear scales}
\label{app:bias_linear}
In \cref{fig:MCMC_relative_bias_linear_ell} we show the relative bias in the same manner as in \cref{fig:MCMC_relative_bias}. Here, however, we excluded non-linear scales from the survey by defining the non-linear scale $R_\mathrm{nl}$ implicitly as:
\begin{equation}
    1 = \int\frac{k^2\mathrm{d}k}{2\pi^2}W^2_{R_\mathrm{nl}}(k)P_\delta(k,\chi)\;,
\end{equation}
where $P_\delta(k)$ is the linear matter power spectrum and $W^2_{R_\mathrm{nl}}$ a suitable weighting function. Via the Limber relation $k = \frac{\ell +0.5}{\chi}$ we define in each tomographic bin a maximum multipole, $\ell_{\mathrm{max},i}$, up to which the corresponding power spectrum is considered for the analysis. In this way we make sure that certain scales do not contribute to the survey. We see that, even when relying on linear scales only, the systematic bias induced can become very big even with a few per cent residual alignments. It should be noted that the dominant contribution on these scales comes from the tidal shearing model for elliptical galaxies. In particular the tidal shearing model can be motivated much more easily since it does not rely on the assumption that the angular momentum of the baryons aligns with the angular momentum of the dark matter halo.

\bibliographystyle{JHEP}
\bibliography{My_Library.bib}

\end{document}